\documentclass[aps,prd,preprint]{revtex4}

\usepackage{longtable}
\usepackage{graphicx}
\usepackage{color}
\usepackage{tikz}
\usepackage{units}
\usetikzlibrary{shapes,arrows}

\begin{document}

\title{Method for estimation of gravitational-wave transient model parameters in frequency-time maps}
\author{M Coughlin}
\affiliation{Department of Physics, Harvard University, Cambridge, MA 02138, USA}
\author{N Christensen}
\affiliation{Physics and Astronomy, Carleton College, Northfield, Minnesota 55057, USA}
\author{J Gair}
\affiliation{Institute of Astronomy, University of Cambridge, Cambridge, CB30HA, United Kingdom}
\author{S Kandhasamy}
\affiliation{Physics and Astronomy, University of Mississippi, University, MS 38677-1848, USA}
\author{E Thrane}
\affiliation{LIGO Laboratory, California Institute of Technology, MS 100-36, Pasadena, CA, 91125, USA}

\begin{abstract}
A common technique for detection of gravitational-wave signals is searching for excess power in frequency-time maps of gravitational-wave detector data. In the event of a detection, model selection and parameter estimation will be performed in order to explore the properties of the source. In this paper, we develop a  Bayesian statistical method for extracting model-dependent parameters from observed gravitational-wave signals in frequency-time maps. We demonstrate the method by recovering the parameters of model gravitational-wave signals added to simulated advanced LIGO noise. We also characterize the performance of the method and discuss prospects for future work.
\end{abstract}

\maketitle

\section{Introduction}
\label{sec:Intro}

The Laser Interferometer Gravitational-wave Observatory (LIGO) \cite{LIGO}, Virgo \cite{VIRGO}, and GEO600 \cite{GEO600} detectors are part of a network of gravitational-wave (GW) detectors seeking to make direct observations of GWs. Previous analyses of the data have included searches targeting the coalescence of neutron stars or black holes \cite{S6Highmass,S6Lowmass}, short-duration bursts \cite{PhysRevD.85.122007}, isolated neutron stars \cite{0004-637X-713-1-671}, and a stochastic background of GWs \cite{S5StochasticNature,PhysRevD.85.122001}. LIGO and Virgo are currently upgrading to Advanced LIGO (aLIGO) and Advanced Virgo (AdV), which will improve their strain sensitivities by one order of magnitude over the strain sensitivites achieved during previous science runs \cite{aLIGO,AdVirgo}. These will be joined by GEO-HF \cite{0264-9381-23-8-S26} and KAGRA \cite{0264-9381-27-8-084004}. To date, none of the above searches have resulted in a GW detection, although with the current upgrades, the chances will increase significantly. In the event of a detection, one can perform model selection and parameter estimation in order to further explore the properties of the sources. Model selection and parameter estimation are topics of great interest in the GW community (see e.g. \cite{MCMCMethods,NestedSampling,MultiNestLISA,MultiNestGW,S6PE,SMEE,BayesianReconstruction}). 

The binary coalescence of compact objects are well-studied sources of GWs, and the most up-to-date models for the waveforms produced in these systems include most of the physical effects that influence the signals, including tidal and spin effects \cite{0264-9381-27-17-173001}. Searches and parameter estimation for these sources rely on matched filtering of the signal seen by detectors using models of the signals. Because models for these sources are thought to be reliable, a full Bayesian analysis utilizing matched filtering is possible for these sources, and the ability to precisely estimate injected waveform parameters for these sources has been demonstrated \cite{S6PE}. 

On the other hand, GW bursts cannot be modeled precisely (by assumption). GW emission by core-collapse supernovae is one such example. A number of competing models for the mechanism that drives the core-collapse exist, and each model produces qualitatively different waveforms. Logue et al. demonstrated that a principal component analysis can be used to determine the correct model of injected GW waveforms by the computation of the Bayesian odds ratio \cite{SMEE}. Principle component analysis has also been used to reconstruct the stellar core-collapse GW signal after finding the amplitude of the individual principle components and arrival times \cite{BayesianReconstruction}.

Parameter estimation of signal models requires, at first, GW detection with high significance. A common technique for detection of GW bursts is searching for excess power in frequency-time (denoted $ft$) maps of GW detector data \cite{X-Pipeline,CoherentWaveBurst,STAMP}. Matched filtering is not used for these signal types because the precise waveforms are unknown. Excess power searches provide an effective alternative to matched filtering for such signals. Some signals can be well-approximated by parameterized spectrogram curves which incorporate the salient features of the signals, and these curves can be used to focus the search with a ``phase-less template bank'' \cite{Stochtrack,Stochsky}. 

In this paper, we present a method for parameter estimation using GW tracks in $ft$-maps. We explore the possibility of performing parameter estimation and model selection, assuming that a search has been performed and a signal detected. We seek to address the question of how to fit the model parameters. As a concrete example, we show the recovery of parameters of an r-mode signal injected into simulated detector data. These GW sources are unstable oscillation modes which dampen the rotation of neutron stars by the emission of GWs \cite{Alford:2011pi}. We show how to estimate parameters such as the r-mode saturation amplitude, which is the amplitude above which the emitting neutron star will collapse into a black hole.

The remainder of this paper is organized as follows. We discuss the methods used to extract waveform parameters from tracks in $ft$-maps in section \ref{sec:ParameterEstimation}. To demonstrate the method and performance of parameter recovery, we perform sample injections into simulated aLIGO colored Gaussian noise and recover their parameters in section \ref{sec:Results}. We conclude with a discussion of topics for further study in section \ref{sec:Conclusion}.

\section{Formalism}
\label{sec:ParameterEstimation}

In this section, we discuss the methods used to extract waveform parameters from tracks in $ft$-maps. To begin, we review the data products used in the detection of unmodeled GW transients.

\subsection{Frequency time maps}

Many searches for GW bursts rely on searching for excess power in $ft$-maps of GW detector data \cite{X-Pipeline,CoherentWaveBurst,STAMP}. The maps are computed by dividing detector strain time series into segments and computing a Fourier transform of each segment. Each column in the map corresponds to one of these segments. Searches for long-duration GW bursts in particular use the cross-correlation of two GW strain channels from spatially separated detectors to construct $ft$-maps of \emph{cross-power} signal-to-noise ratio, $\rho(t;f)$ \cite{STAMP}.

\begin{equation}
  \rho(t;f)\equiv\hat{Y}(t;f)/\hat\sigma(t;f) .
  \label{eq:rho}
\end{equation}
where $t$ is the time of the segment, $f$ is the frequency, $\hat{Y}(t;f)$ is an unbiased estimator for GW power and $\hat\sigma^2(t;f)$ is its variance. Arrays of $\rho(t;f)$ are visualized as $ft$-maps.

\begin{figure*}[t]
 \includegraphics[width=2.9in]{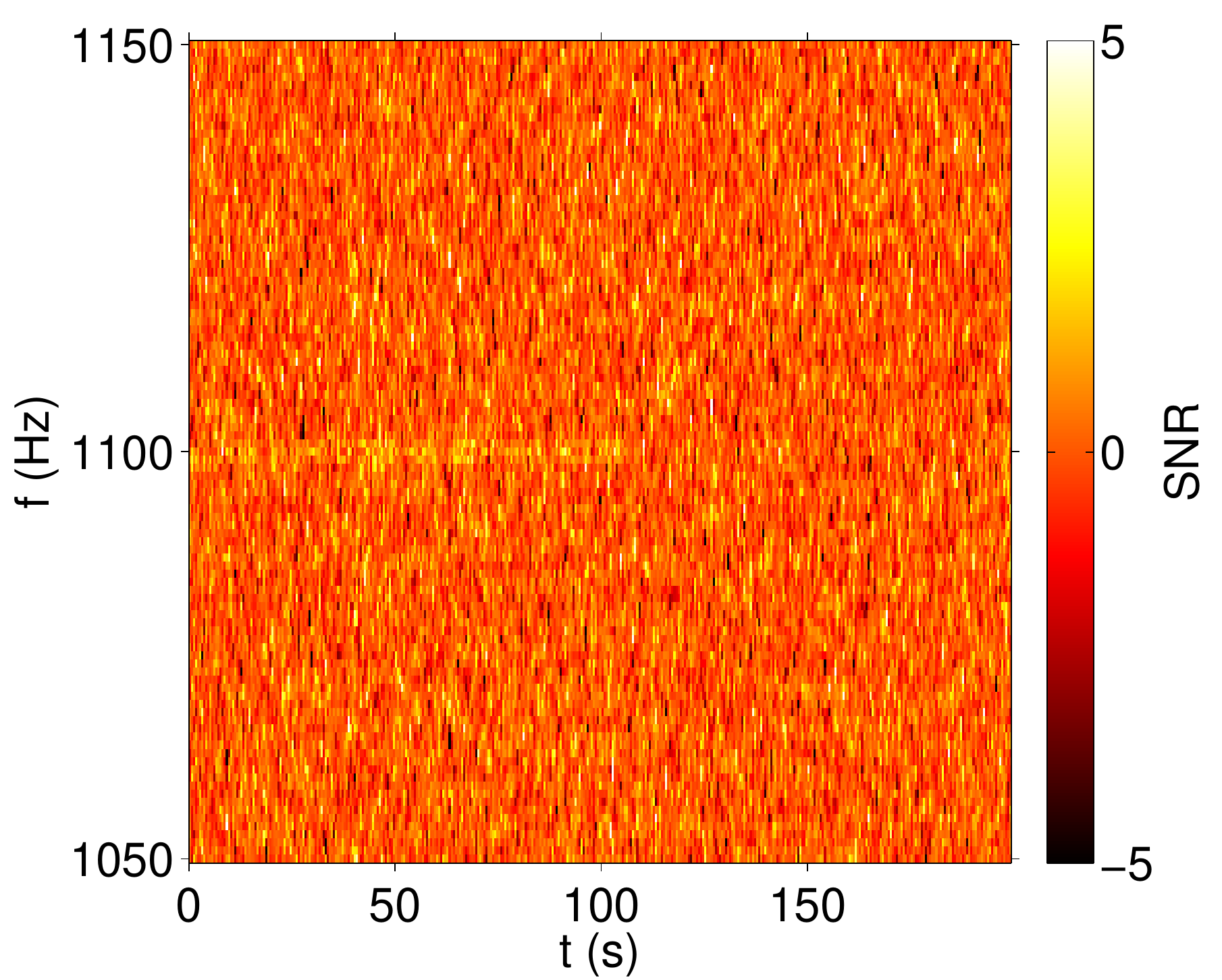}
 \includegraphics[width=2.9in]{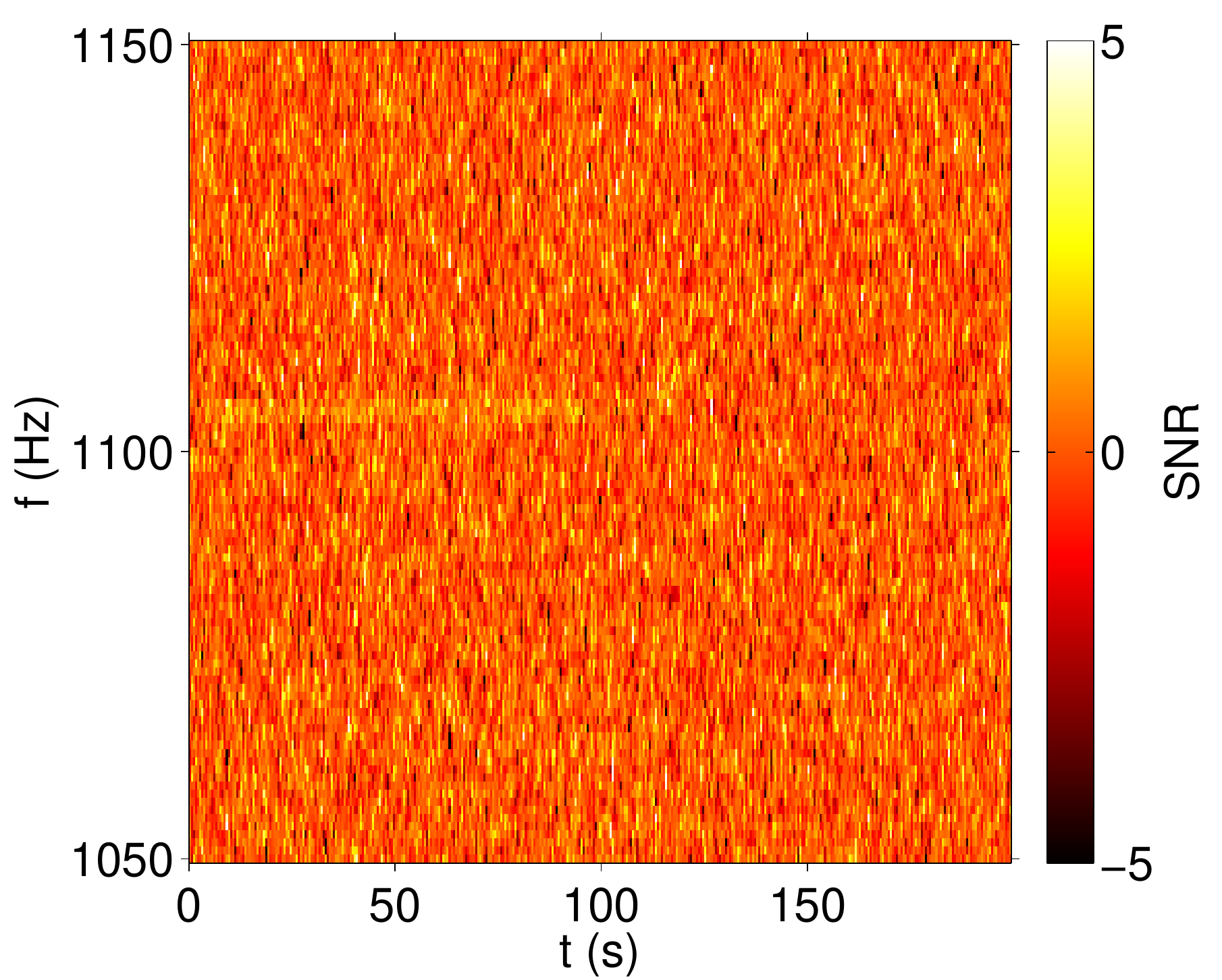}\\
 \includegraphics[width=2.9in]{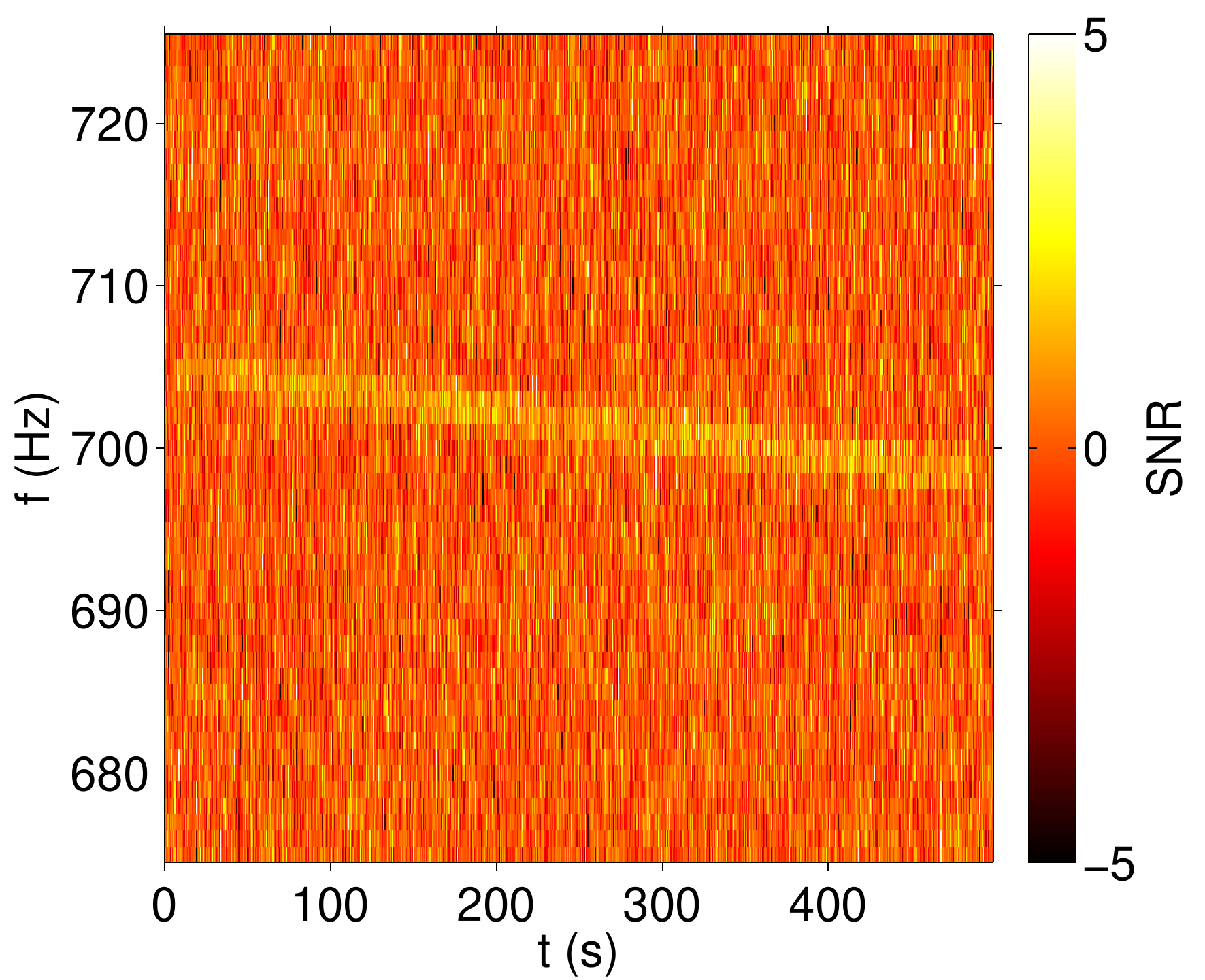}
 \includegraphics[width=2.9in]{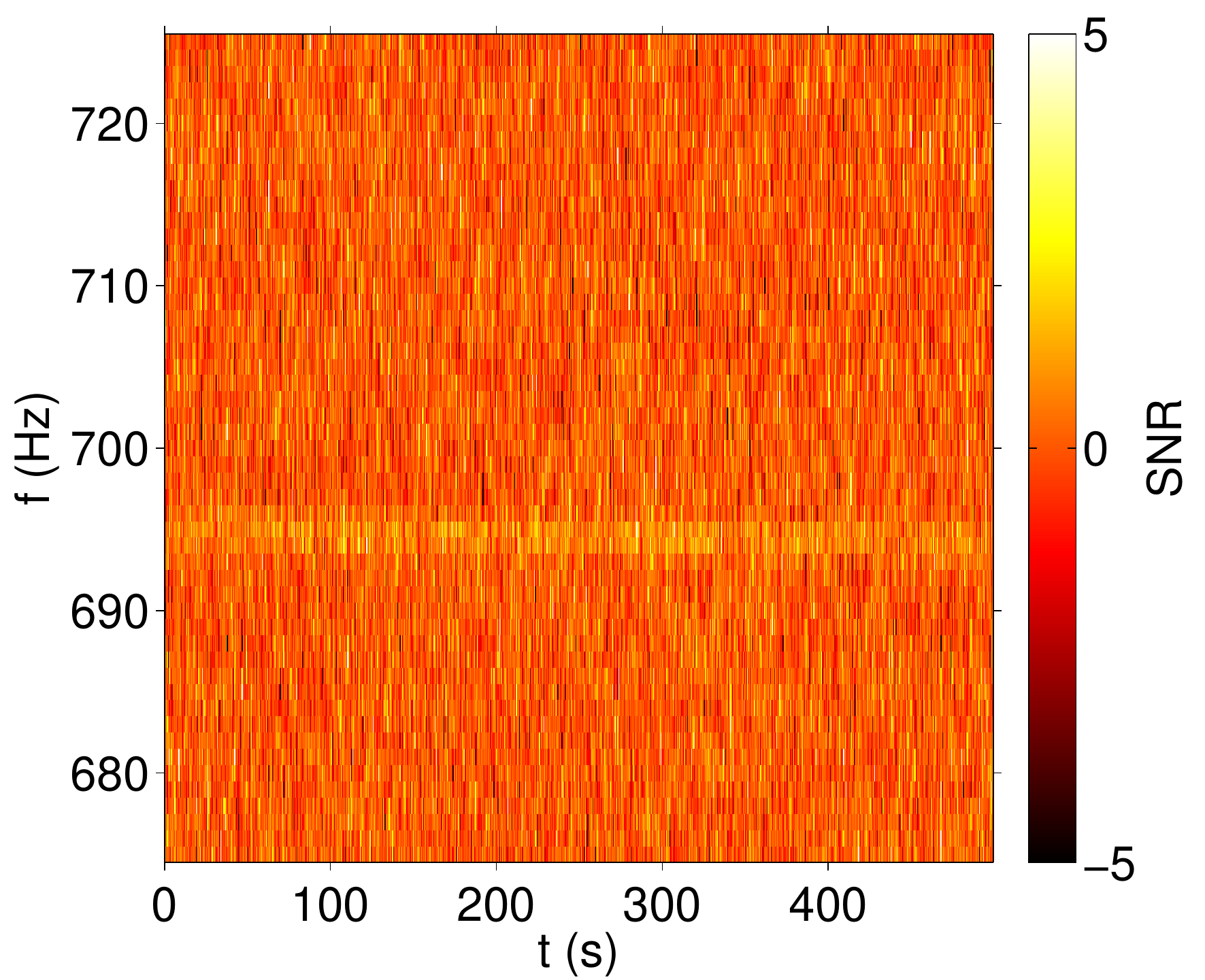}\\
 \caption{$ft$-maps of $\rho(t;f)$ with injected signals. The top row consists of sine-Gaussian injections \cite{PhysRevD.85.122007}; on the left is an injection with $f_0$ = 1100\,Hz and $\tau$ = 100\,s and on the right is $f_0$ = 1095\,Hz and $\tau$ = 80\,s. The bottom row consists of r-mode injections \cite{PhysRevD.58.084020}; on the left is an injection with $f_0$ of 705\,Hz and a saturation amplitude $\alpha$ = 0.3, and on the right is an injection with $f_0$ of 695\,Hz and a saturation amplitude $\alpha$ = 0.1. The injections are performed at a distance at which a GW signal can be observed above threshold with false alarm probability = 0.1\% and false dismissal probability = 50\% using a seed-based clustering algorithm \cite{burstegard}. This corresponds to a matched filter SNR of about 20 for the sine-Gaussian injections and about 30 for the r-modes \cite{PhysRevD.85.042001}.}
 \label{fig:RModesInjectionsSNR}
\end{figure*}

GWs appear as tracks or blobs on $ft$-maps. The morphology of the GW track depends on the source. If the signal is sufficiently loud, compact binaries appear as chirps of increasing frequency, while continuous-wave isolated neutron star sources appear as narrowband, horizontal lines. Fig.~\ref{fig:RModesInjectionsSNR} shows $ft$-maps of example sine-Gaussian injections with different durations (top row) and r-mode injections with different saturation amplitudes (bottom row). 

Given an $ft$-map, GW searches employ pattern recognition algorithms to identify potentially significant clusters of pixels \cite{STAMP}. Next, the pattern-recognition algorithms are run repeatedly on noise-only maps to generate background statistics. These noise-only maps are created using GW detector strain data with a time-shift that removes any potential GW signal. Using time shifts to study noise and injections to study detection efficiency, false alarm and false dismissal rates can be estimated, and detections can potentially be made.

\subsection{Waveform models}
A metric perturbation, $h_{ab}$, can be written as a linear combination of two polarizations, $h_+$ and $h_\times$.

\begin{equation}
h_{ab} = \left(
\begin{array}{ccc}
h_+ & h_\times \\
h_\times & -h_+
\end{array}
\right)
\end{equation}
Far from an elliptically polarized source, we can write the metric pertubation as

\begin{equation}
h_+(t) = h_{\mathrm{amp}}(t) (1 + \cos^2 \iota) \cos \psi(t)
\label{eq:hplus}
\end{equation}
and

\begin{equation}
h_{\times}(t) = 2 h_{\mathrm{amp}}(t) \cos \iota \sin \psi(t),
\label{eq:hcross}
\end{equation}
where $h_{\mathrm{amp}}$ is the strain amplitude, $\iota$ is the inclination angle and $\psi$ is the polarization angle. In the analysis below, we assume that we have a face-on source, so $\iota = 0$. We also assume that $\psi = 0$. This is for the sake of simplicity; in theory, one may estimate $\iota$ and $\psi$ as well, but an analysis involving these parameters is beyond the scope of this paper.

GW detectors measure strain, $h_0(t)$,

\begin{equation}
h_0(t) = h_+(t) F_+(t) + h_\times(t) F_\times(t)
\end{equation}
where $F_+(t)$ and $F_\times(t)$ are the detector antenna response functions to the two polarizations \cite{lrr-2009-2}. GW amplitudes are sometimes characterized by the root-sum-square amplitude,$h_\textrm{rss}$, defined as
\begin{equation}
h^2_{\textrm{rss}} = \int [h_+^2(t) + h_\times^2(t) ] dt.
\end{equation}

Each GW burst creates a specific pattern in $ft$-maps which depends on astrophysical parameters. In this study, three different models of GW signals are used. The first is a sine-Gaussian, which is commonly used in searches for GW bursts \cite{PhysRevD.85.122007}. This model depends on four parameters: the waveform duration, $\tau$, the start frequency, $f_0$, the signal distance, $D$, and the time of the maximum of burst, $t_0$. It has the following form

\begin{eqnarray}
h_{0}(t) = k \frac{\mathrm{exp}(- (\frac{(t-t_0)^2}{4 \tau^2} + 2 \pi i (t-t_0) f_0) )}{ D \left(2 \pi \tau^2 \right) ^{1/4}}, \\
h_+ (t) = \mathrm{Re}[h_0(t)], \hspace{10pt} h_\times (t) = \mathrm{Im}[h_0(t)].
\end{eqnarray}
where $k$ is a constant. The second waveform represents a simple r-mode model, based on a model by Owen et al. \cite{PhysRevD.58.084020}. This model depends on three parameters: the saturation amplitude, $\alpha$, the start frequency, $f_0$, and the signal distance, $D$. It has the following form 

\begin{eqnarray}
f(t) = \left(\frac{1}{f_{0}^{-6} - 6 k t}\right)^{1/6}, \\
h_+ (t) = h_0 \cos (2 \pi f(t) t), \hspace{10pt} h_\times (t) = h_0 \sin (2 \pi f(t) t)
\label{eq:RMode}
\end{eqnarray}
where
\begin{eqnarray}
h_{0} = 3.6 \times 10^{-23} \alpha \left(\frac{f(t)}{1000}\right)^3 / D, \\
k = - 1.8 \times 10^{-21} \alpha^2 \textrm{Hz}^{-5}.
\end{eqnarray}

The third waveform is a slowly varying sinusoid waveform with a time-varying frequency, $f = f_0 + \dot{f} t$. This model is chosen here as its morphology is similar to the r-modes. This model depends on three parameters: the time derivative of signal frequency, $\dot{f}$, the start frequency, $f_0$, and the signal distance, $D \propto 1/h_\textrm{rss}$.

\begin{equation}
h_{0}(t) = c \frac{\mathrm{exp}(2 \pi i (f_{0} + \dot{f} t) t)}{D}.
\label{eq:CW}
\end{equation}
where $c$ is a constant. $h_+$ and $h_\times$ are calculated in the same way as the sine-Gaussian.

\subsection{Likelihood}

We use the above models to illustrate our method. Fig.~\ref{fig:RModesInjectionsSNR} shows two pairs of $ft$-maps of cross-power with sine-Gaussian and r-mode injections. Our goal is to determine, based on the map structure, the parameters which best fit the models. In order to estimate the parameters, we employ a likelihood formalism.

The first step is to compute the probability distribution of $\rho_B(t;f)$ due to background, $f_B(\rho)$. A distribution valid for Gaussian and stationary noise is derived in \ref{sec:STAMPDist}. In cases for which an analytic distribution is impractical to construct, it can be estimated from time-shifted data. In the analysis below, we assume Gaussian noise for simplicity; the use of time-shifted distributions will be explored in a future study. We also assume that $\rho_B(t;f)$ of each pixel is drawn from the same distribution. The second step is to determine the contribution to $\rho$ from a signal. We denote the expected $\rho$ value due to a signal with parameters $\theta$ by $\rho_S(\theta)$. We calculate the expected contribution using an approximation described in \ref{sec:ErrorApproximation}. The assumption leads to an approximate formulation, which can be made more accurate by performing injection studies and computing the distributions with signals present.
 

Armed with both the distribution of $\rho_B(t;f)$ due to background as well as that of the waveform models, $\rho_S(t;f)$, we are able to construct our likelihood. The idea is to subtract $\rho_S(\theta)$ from $\rho$, which would just leave detector noise if $\rho_S(\theta)$ was the correct waveform model. Minimizing the residuals maximizes the likelihood function. The probability density function describing the residuals is $p(\rho-\rho_S(\theta))$, which is calculated by finding the probability that $\rho-\rho_S(\theta)$ is due to noise, as given by $f_B(\rho)$. The likelihood is

\begin{equation}
L(\{\rho_i\}|\theta) = \Pi_{i=1}^{N} p(\rho_i - \rho_{s_i}(\theta)|\theta)
\label{eq:Likelihood}
\end{equation}
where $i$ is the pixel index and $N$ is the number of pixels in the $ft$-map. The goal is to maximize the likelihood in order to determine confidence intervals for $\theta$. 

Ideally, one would produce an $ft$-map of the GW signal and evaluate Eq.~(\ref{eq:Likelihood}) for every set of possible parameters. In this way, we could generate the posterior density functions (PDFs) for the relevant model parameters. Because this is computationally intractable, we use algorithms that efficiently sample the posterior while minimizing the computational burden. In the examples below, we use flat, non-informative priors on the parameters of interest. This could be modified, for example, in the event of an r-mode detection, where models predict a small value of $\alpha$.

There are three main algorithms presently used to rapidly evaluate the posterior in GW parameter estimation and model selection: Markov Chain Monte Carlo (MCMC) \cite{S6PE,MCMCMethods}, Nested Sampling \cite{Skilling,NestedSampling}, and MultiNest \cite{MultiNest,MultiNestLISA,MultiNestGW}. Nested Sampling and MultiNest calculate the Bayesian evidence for a given set of parameters, which can be used to assign relative probabilities to different models. We use a MATLAB implementation of Nested Sampling and MultiNest \cite{MatlabMultiNest}, which implements the MultiNest algorithm, as described in \cite{MultiNest}, and Nested Sampling, as described in \cite{Skilling}.

\section{Demonstration}
\label{sec:Results}

In this section, we present two examples of parameter estimation of toy model waveforms. We inject GW signals into simulated aLIGO colored Gaussian noise and create $ft$-maps based on the resulting timeseries. We use the design sensitivity aLIGO noise curve \cite{aLIGO}. We perform injections at the waveform models' detection distance, which we define as the distance at which a GW signal can be observed above threshold with false alarm probability = 0.1\% and false dismissal probability = 50\% using a seed-based clustering algorithm \cite{burstegard}. In order to construct parameter posterior distributions, we produce $ft$-maps containing only GW signals for various sets of parameters. Eq.~(\ref{eq:Likelihood}) is evaluated repeatedly for each set of parameters. From the equation, the likelihood is maximized for those parameters that best minimize the residuals. Parameter posterior distributions are constructed for
parameter sets of equal likelihood that maximize this likelihood.

\subsection{Sine-Gaussian Burst}

\begin{figure*}[t]
 \includegraphics[width=2.9in]{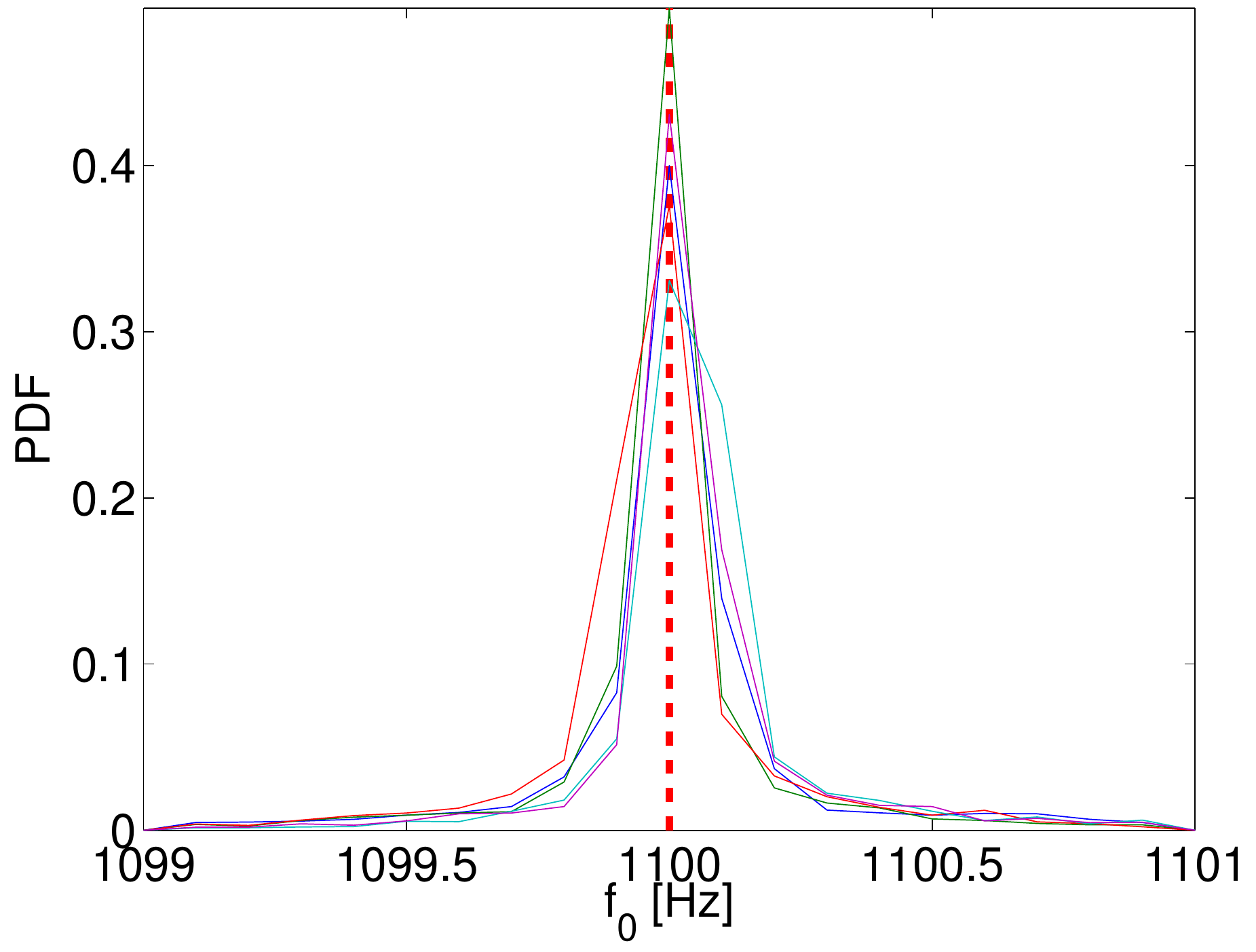}
 \includegraphics[width=2.9in]{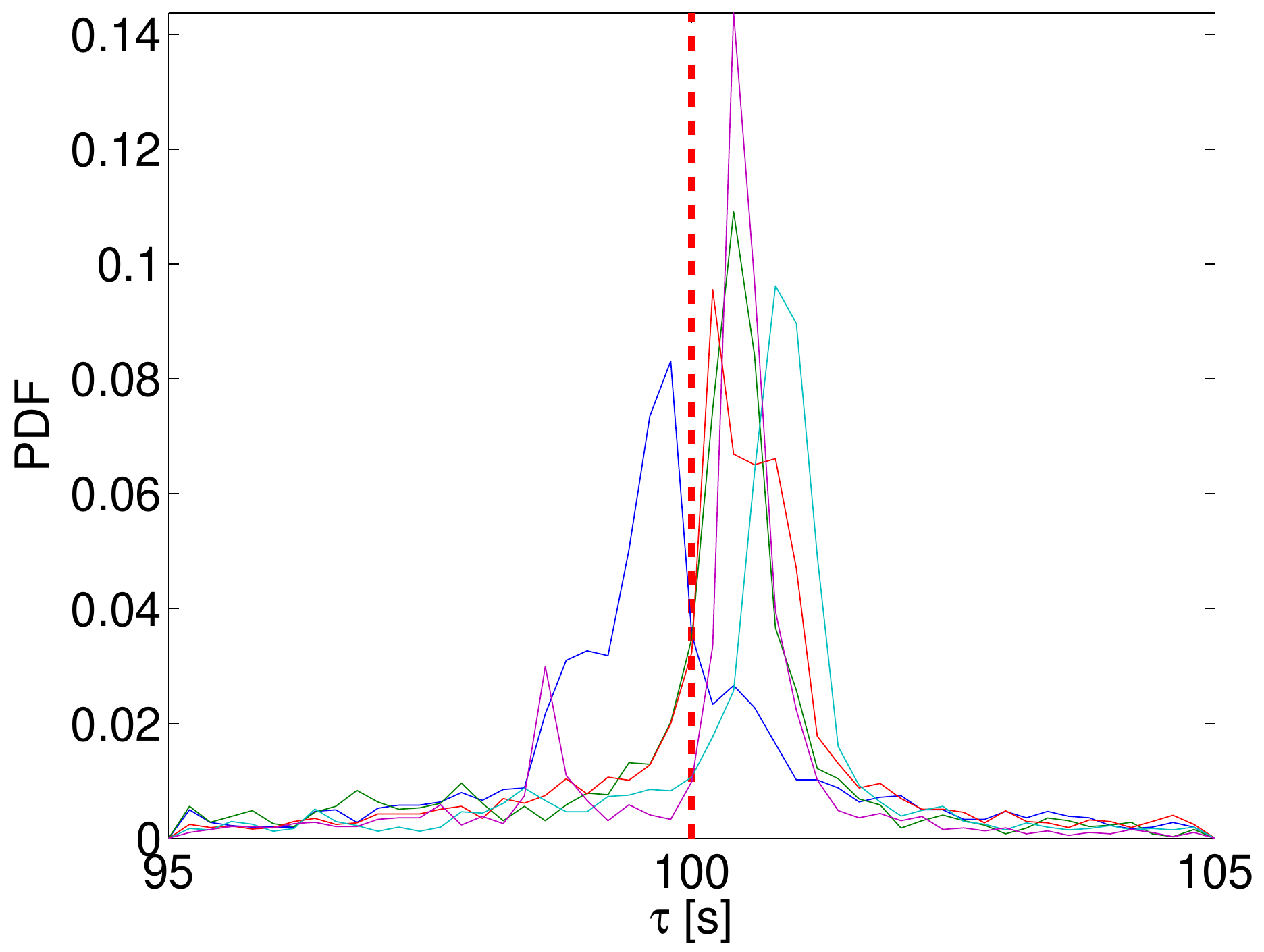}\\
 \includegraphics[width=2.9in]{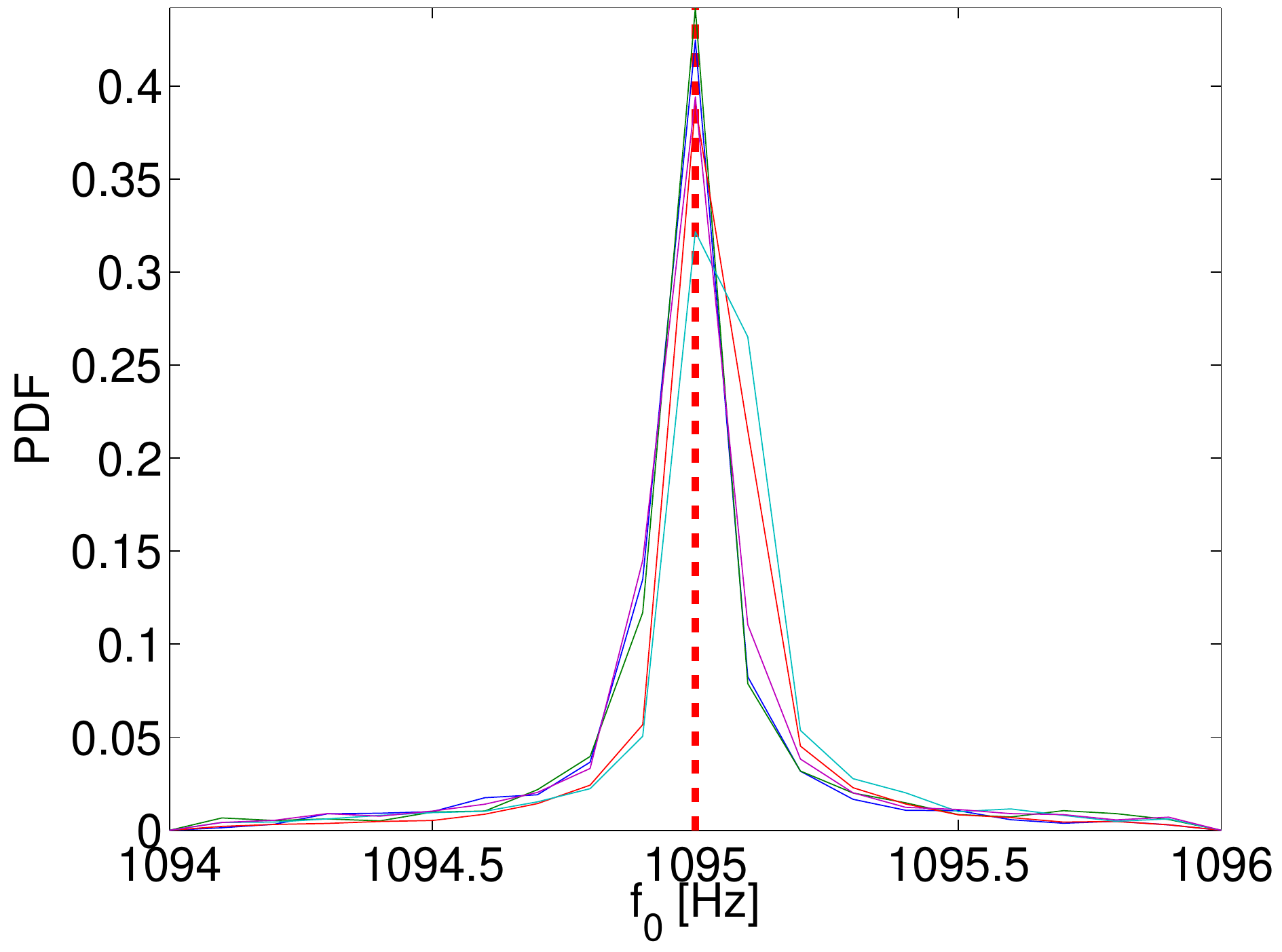}
 \includegraphics[width=2.9in]{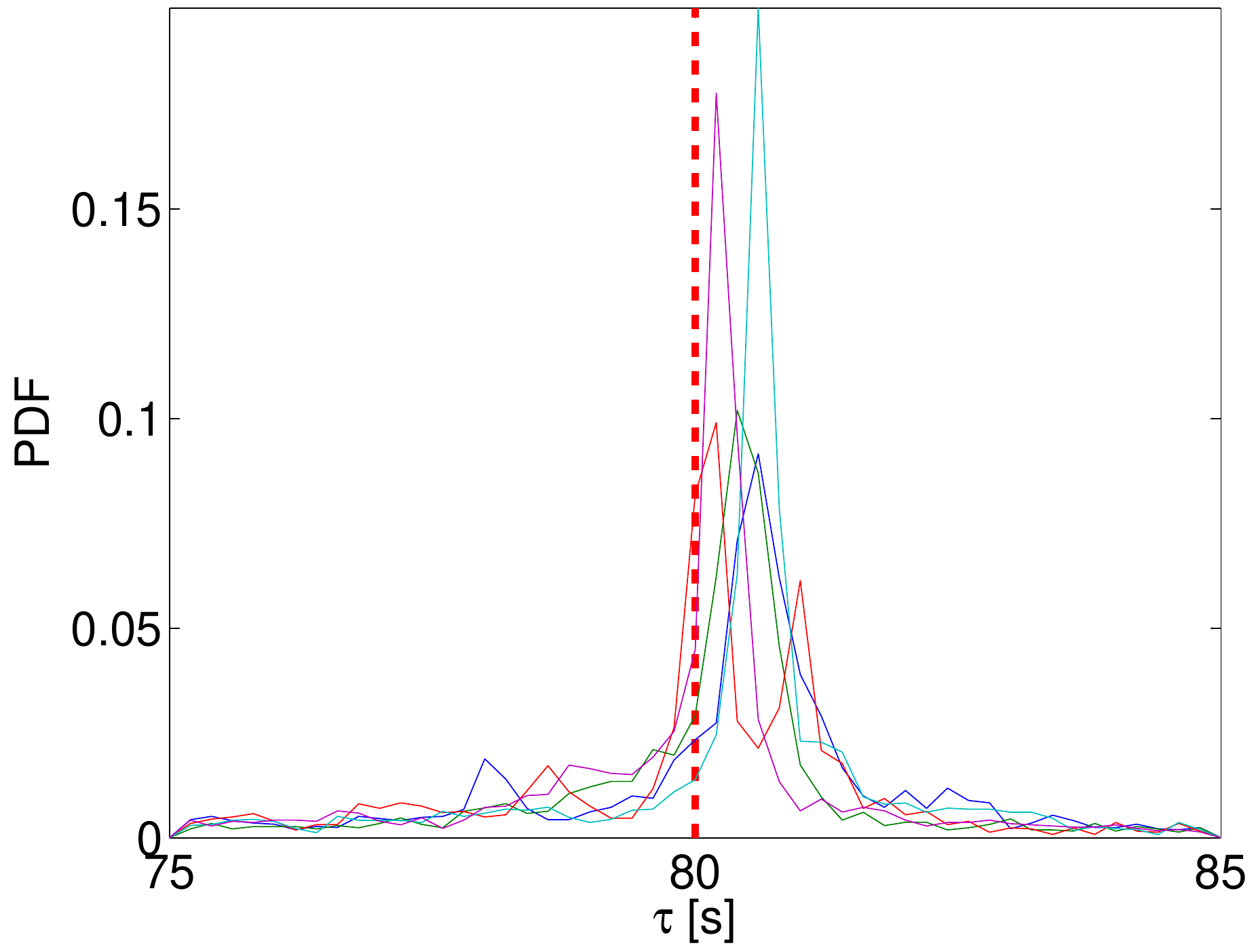}\\
 \caption{PDFs of distributions for injected sine-Gaussian signals. Each injection is performed into multiple simulated aLIGO colored Gaussian noise realizations. The injections are performed at a distance at which a GW signal can be observed above threshold with FAP = 0.1\% and FDP = 50\%. This corresponds to a matched filter SNR of about 30 for these injections. The PDF for each injection is plotted in a different color. The red dotted line shows the true injected value. The plot on the left is the PDF of $f_0$. The plot on the right is the PDF of $\tau$. The top row corresponds to an injection of $f_0$ = 1100\,Hz and $\tau$ = 100\,s. The top row corresponds to an injection of $f_0$ = 1095\,Hz and $\tau$ = 80\,s. The uncertainty in $f_0$ is negligible, while the $\tau$ sampling is within tens of percent.}
 \label{fig:ChirpletInjectionsSamples}
\end{figure*}


Fig. \ref{fig:ChirpletInjectionsSamples} shows the distribution of posterior samples for both $f_0$ and $\tau$ for one of the injections. The posteriors of the parameters are consistent with the injected values. In general, the recoveries for $f_0$ are within a few percent of the true value for all injections. The recoveries for $\tau$ are within tens of percent.  

\subsection{R-mode}


\begin{figure*}[t]
 \includegraphics[width=2.9in]{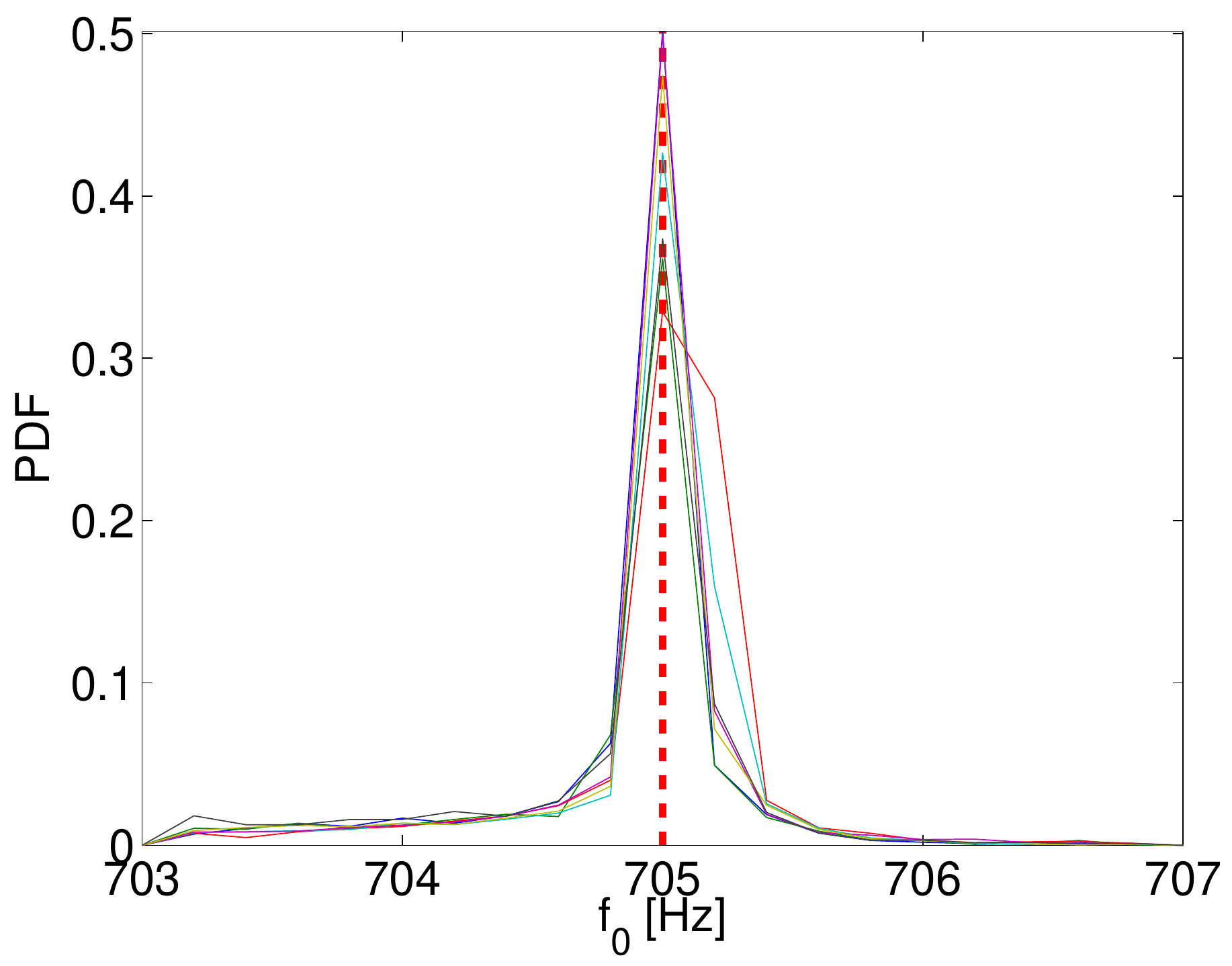}
 \includegraphics[width=2.9in]{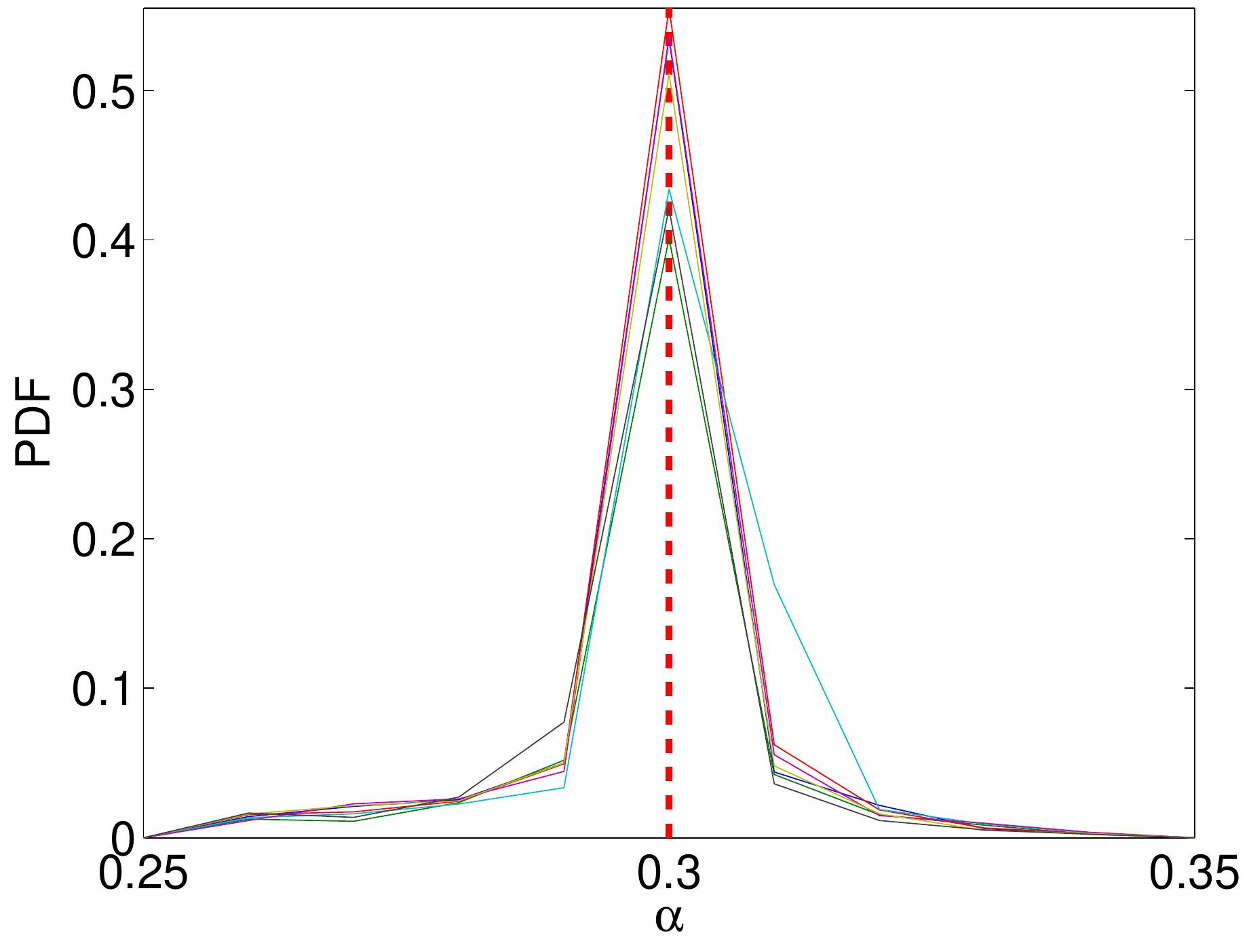}\\
 \includegraphics[width=2.9in]{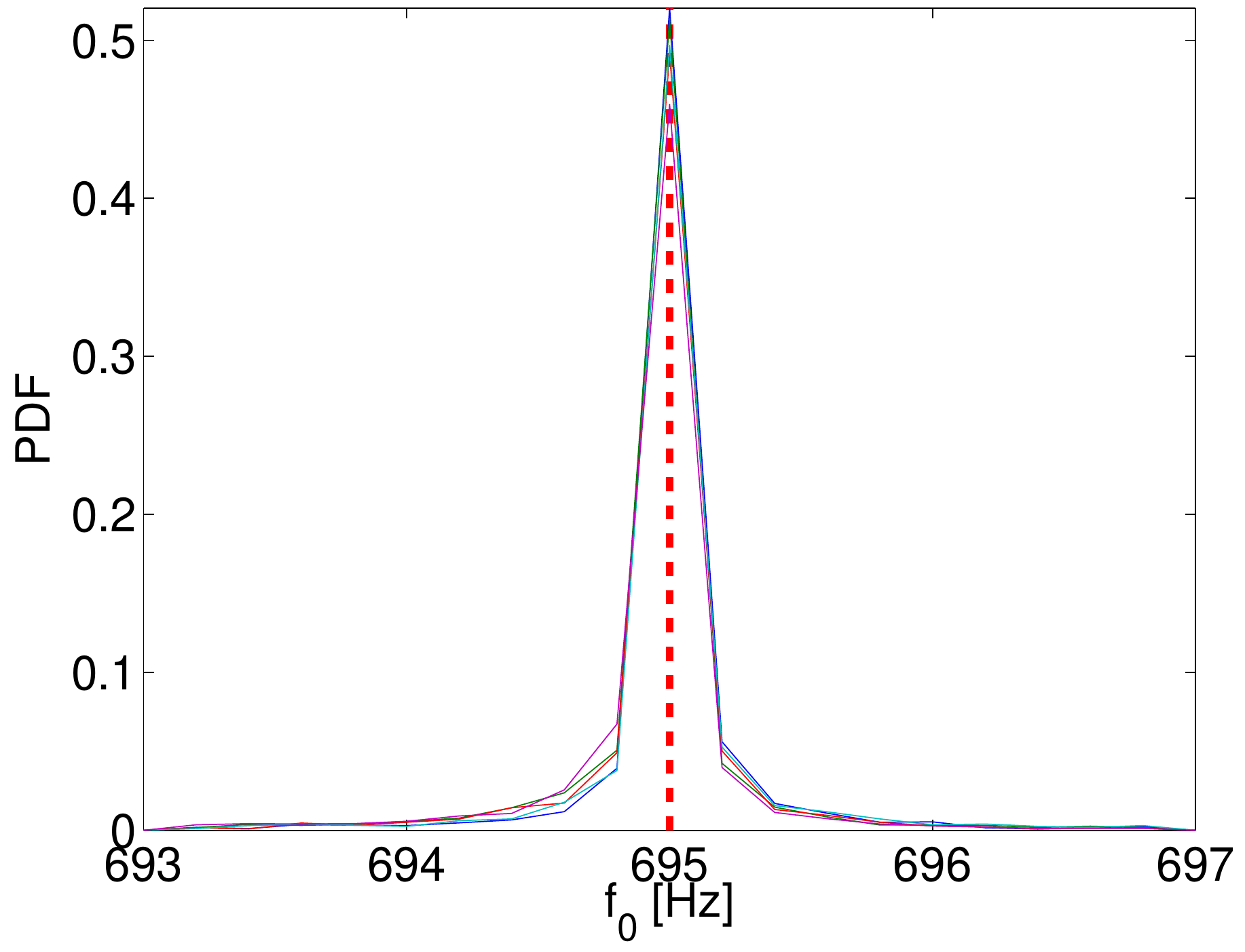}
 \includegraphics[width=2.9in]{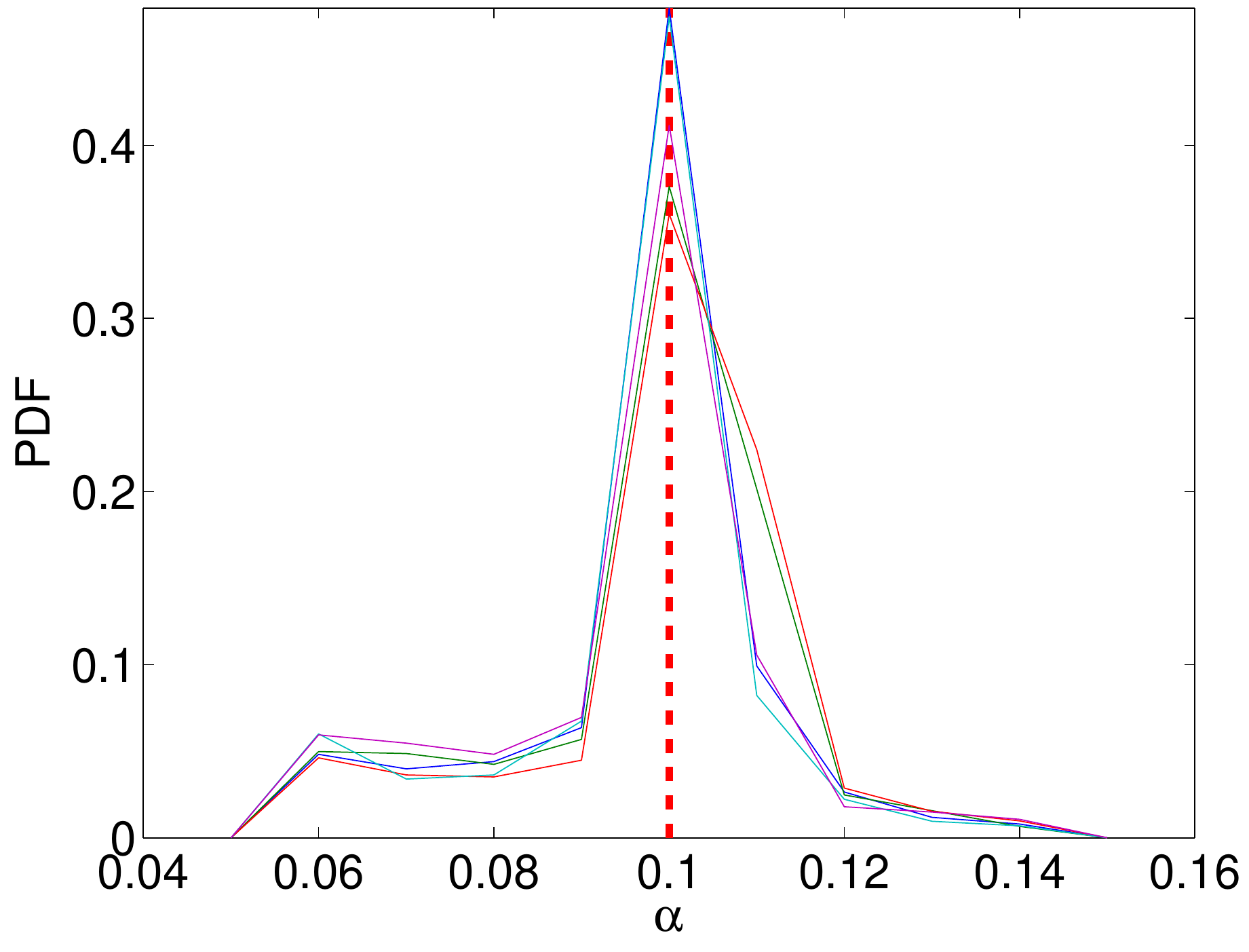}\\
 \caption{PDFs of distributions for injected r-modes signals. Each injection is performed into multiple simulated aLIGO colored Gaussian noise realizations. The injections are performed at a distance at which a GW signal can be observed above threshold with FAP = 0.1\% and FDP = 50\%. This corresponds to a matched filter SNR of about 20 for these injections. The PDF for each injection is plotted in a different color. The red dotted line shows the true injected value. The plot on the left is the PDF of $f_0$. The plot on the right is the PDF of $\alpha$. The top row corresponds to an injection of $f_0$ = 705\,Hz and $\alpha$ = 0.3, while the bottom row is an injection of $f_0$ = 695\,Hz and $\alpha$ = 0.1. }
 \label{fig:RModesInjectionsSamples}
\end{figure*}

We perform injections of the r-mode waveforms with $f_0$ = 705\,Hz and $\alpha$ = 0.3, as well as an injection of $f_0$ = 695\,Hz and $\alpha$ = 0.1. Fig.~\ref{fig:RModesInjectionsSamples} shows the performance of the parameter recoveries. In general, the recoveries for $\alpha$ and $f_0$ are within a few percent for all injections. We can ask whether the r-mode or the varying sinusoidal model is a better description of the $ft$-map for the r-mode signal. This can be done by evaluating the Bayes factor, which is the ratio of the evidences for the two models. The evidence computed by the search algorithm for the r-mode model was $1.23 \times 10^4$, while the evidence for the CW model was $1.15 \times 10^4$, meaning the Bayes factor is 700. This implies that the r-mode model is strongly favored over the varying sinusoidal model. This is despite the fact that the varying sinusoid is a good fit for the linear portion of the r-mode parameter space.
 
\begin{figure*}[t]
 \includegraphics[width=2.9in]{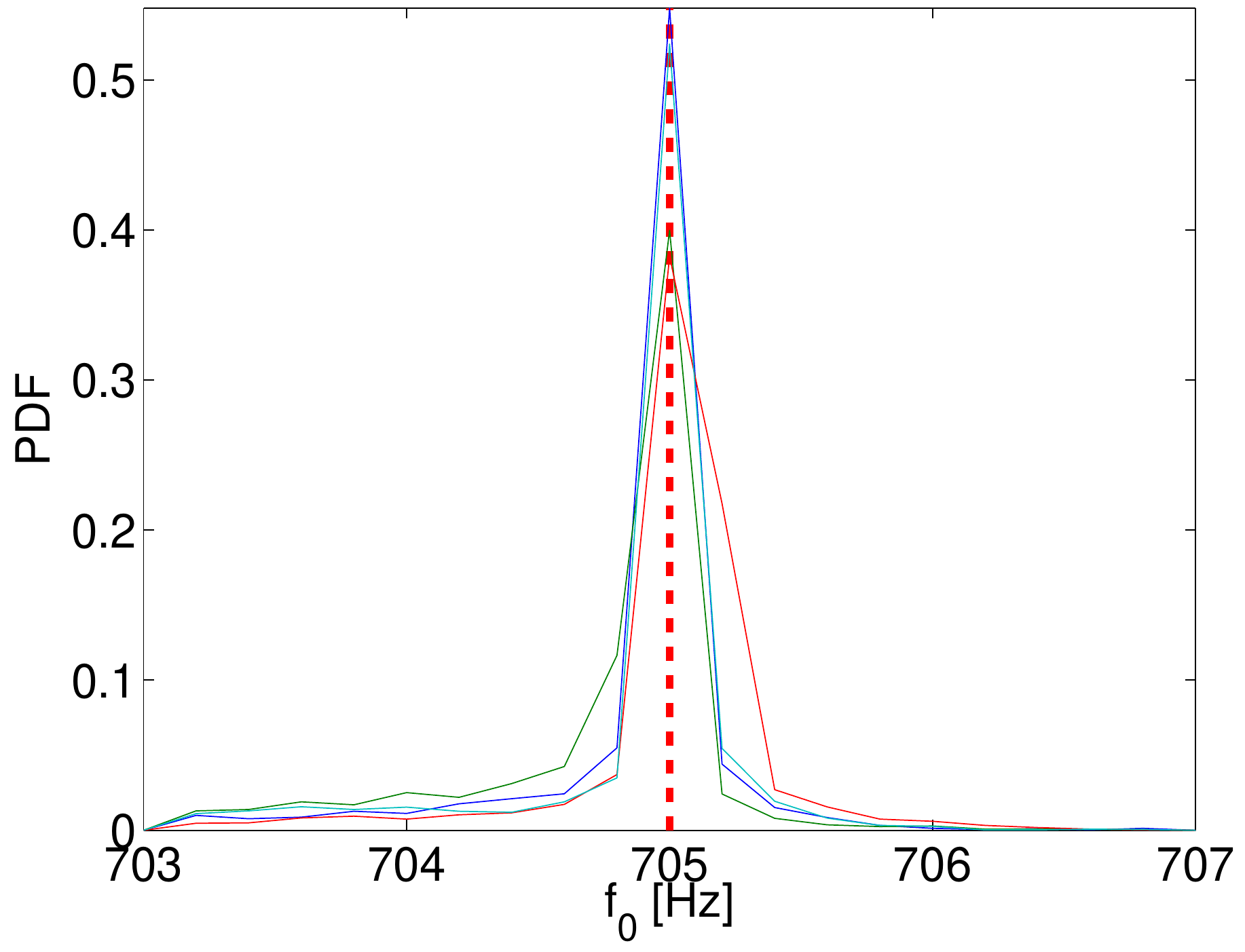}
 \includegraphics[width=2.9in]{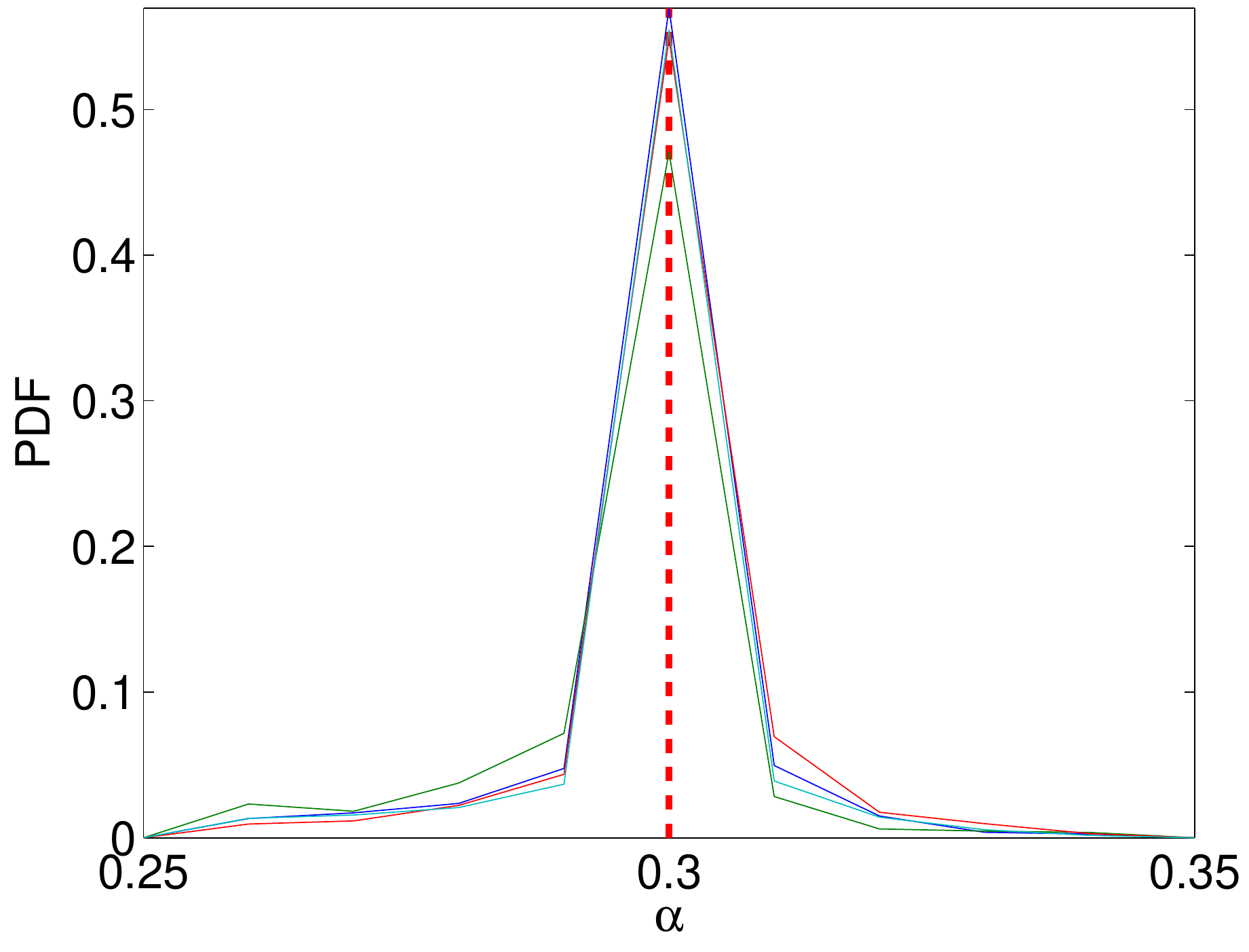}\\
 \includegraphics[width=2.9in]{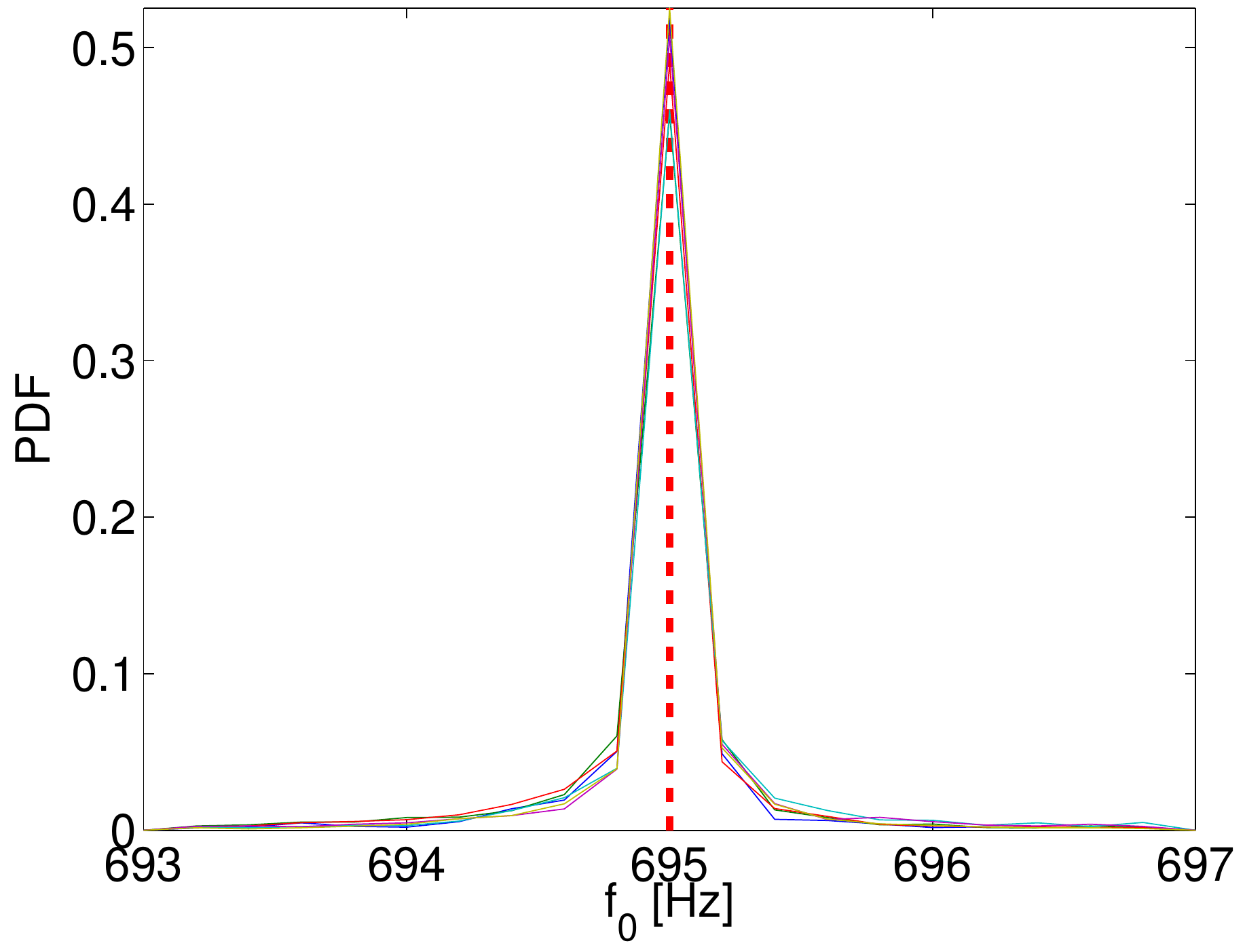}
 \includegraphics[width=2.9in]{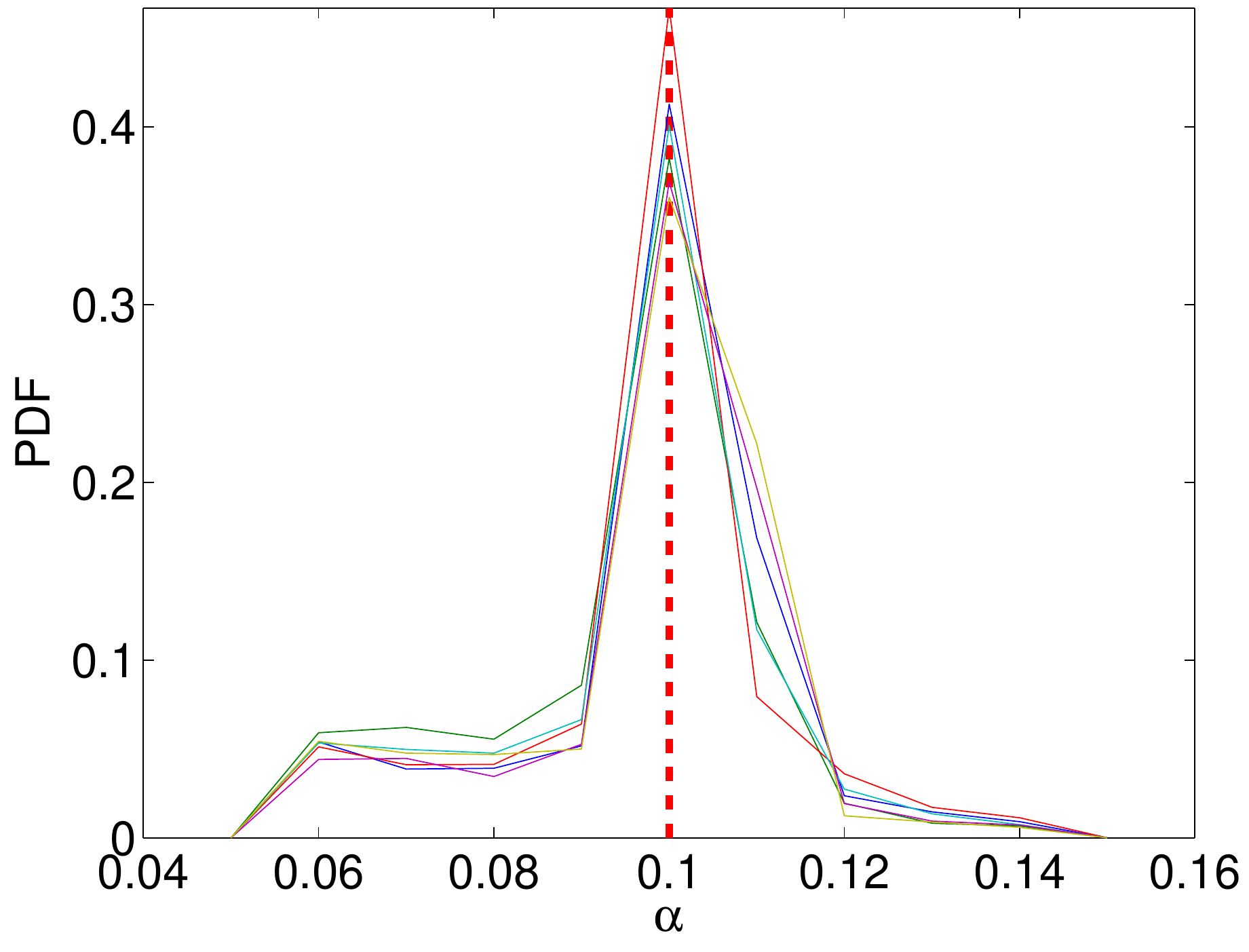}\\
 \caption{Same as Fig.~\ref{fig:RModesInjectionsSamples} for initial LIGO noise recolored to advanced LIGO noise. The parameter recoveries are similar to the Gaussian noise case.}
 \label{fig:RModesInjectionsSamplesRecolored}
\end{figure*}

It is a well-documented fact that in the parameter estimation of compact binary coalescences non-Gaussian noise can significantly affect the posterior recoveries \cite{PhysRevD.88.084044}. Therefore, it is worthwhile to test the algorithm when the noise background is non-Gaussian and non-stationary and thus violates the approximations that go into deriving the noise model used in this analysis. For this reason, we repeat the test with initial LIGO noise which has been recolored to match the design sensitivity of Advanced LIGO \cite{LIGO,aLIGO}. We introduce an artificial time-shift in the initial LIGO data to remove any potential GW signals present ~\footnote{The data are taken in between GPS times 822917487 and 847549782.}. Fig.~\ref{fig:RModesInjectionsSamplesRecolored} shows the performance of the recoveries for injections into the recolored noise, using the noise model that assumes Gaussian noise. We find that the performance is similar to that of the Gaussian noise case. Because the data segments analyzed and waveforms are long, it is likely that our pipeline is less susceptible to noise transients than the compact binary case.

\section{Conclusion}
\label{sec:Conclusion}

In this paper, we have demonstrated the ability to perform basic parameter estimation on GW signals from their signature in $ft$-maps. We described the likelihood method used and showed that these methods correctly recovered the parameters of waveform models and were able to differentiate between two similar models. 

In the future, we will move beyond the generic models presented here to more complicated models. This will be necessary to identify the physics underlying a particular GW source by distinguishing between different variations of similar models. Also, further studies of the assumptions made in the paper will be conducted. We have assumed that the noise is Gaussian and stationary and ignore the correlation between pixels in the maps. We have also assumed that the cross terms when multiplying the noise and waveform signals are zero. There are three complicating assumptions involved in the use of the likelihood here. The first is that in reality the cross-terms of the noise and signal are non-zero (see \ref{sec:ErrorApproximation}). The second is that the cross-power statistic uses adjacent PSD's for the purpose of estimating $\sigma$ (from Eq.~(\ref{eq:rho})), meaning there is a correlation between adjacent pixels (see \ref{sec:STAMPDist}). As such, the multiplication of the pixel probabilities in the $ft$-map, which requires that the probabilities are all independent if one wants a true cumulative probability, is not valid. The third is that real detectors have noise transients and non-stationary noise, which violate some of the approximations used here. One way to rectify this is to perform many injections and measure $f_S(\theta)$ empirically (this distribution would change for each signal model). These issues will be explored in the future.

The use of $ft$-maps to perform parameter estimation has the natural advantage over matched filtering in terms of the speed at which it can be done. Because we fit the amplitude of the waveform to the track in the $ft$-map (removing the phase information), it also means that our signal models do not need to be quite as exact as for parameter estimation relying on matched filtering. It is easier to match the amplitude of the signal than the phase, which is required by matched filtering. We can estimate the potential performance of matched filtering parameter estimation using the Fisher Information Matrix (FIM), which is a tool that has been used to estimate the potential accuracy of parameter estimates for GW signals \cite{PhysRevD.46.5236,PhysRevD.49.2658}. In the limit of high signal-to-noise ratio, the inverse of the Fisher Information Matrix is the variance-covariance matrix of the estimated signal parameters. It provides a first-order estimate of the errors when measuring parameters. Applying this technique to the r-mode model discussed above, this technique finds that the errors would be of order 0.1\% for $f_0$ and 1\% for $\alpha$, which is about an order of magnitude better than for the $ft$-map based technique. This seems reasonable as the matched filtering technique includes phase information and includes none of the approximations. We also expect this to be a reasonable estimate for the signals used above because of the high matched filtering SNRs. This is important as it has been previously shown that the Fisher Information Matrix is biased for near-threshold SNR signals \cite{PhysRevD.88.084013}. Further study may include a detailed comparison between matched filtering and $ft$-map perform parameter estimation.

\section*{Acknowledgments}

The authors would like to thank Matthew Pitkin for help running the likelihood sampler.
MC was supported by the National Science Foundation Graduate Research Fellowship Program, under NSF grant number DGE 1144152.
ET is a member of the LIGO Laboratory, supported by funding from United States National Science Foundation.
NC's work was supported by NSF grant PHY-1204371.
JG's work is supported by the Royal Society.
LIGO was constructed by the California Institute of Technology and Massachusetts Institute of Technology with funding from the National Science Foundation and operates under cooperative agreement PHY-0757058.
This paper has been assigned LIGO document number LIGO-P1400043.

\section*{References}
\bibliographystyle{iopart-num}
\bibliography{references}

\providecommand{\newblock}{}
\begin{thebibliography}{10}
\expandafter\ifx\csname url\endcsname\relax
  \def\url#1{{\tt #1}}\fi
\expandafter\ifx\csname urlprefix\endcsname\relax\def\urlprefix{URL }\fi
\providecommand{\eprint}[2][]{\url{#2}}

\bibitem{LIGO}
{Abbott B et al} (LIGO Scientific Collaboration) 2009 {\em Reports on Progress
  in Physics\/} {\bf 72} 076901

\bibitem{VIRGO}
et~al T~A 2012 {\em Journal of Instrumentation\/} {\bf 7} P03012

\bibitem{GEO600}
{Grote H for the LIGO Scientific Collaboration} 2010 {\em Class. Quantum
  Grav.\/} {\bf 27} 084003

\bibitem{S6Highmass}
{Aasi J et al} (LIGO Scientific Collaboration and Virgo Collaboration) 2013
  {\em Phys. Rev. D\/} {\bf 87}(2) 022002

\bibitem{S6Lowmass}
{Abadie J et al} (LIGO Scientific Collaboration and Virgo Collaboration) 2012
  {\em Physical Review D\/} {\bf 85} 082002
  \urlprefix\url{http://arxiv.org/abs/1111.7314}

\bibitem{PhysRevD.85.122007}
{Abadie J et al} (The LIGO Scientific Collaboration and The Virgo
  Collaboration) 2012 {\em Phys. Rev. D\/} {\bf 85}(12) 122007

\bibitem{0004-637X-713-1-671}
{Abbott B et al} (The LIGO Scientific Collaboration and The Virgo
  Collaboration) 2010 {\em The Astrophysical Journal\/} {\bf 713} 671

\bibitem{S5StochasticNature}
{Abbott B et al} (LIGO Scientific Collaboration and Virgo Collaboration) 2009
  {\em Nature\/} {\bf 460} 990--994

\bibitem{PhysRevD.85.122001}
{Abadie J et al} (LIGO Scientific Collaboration and Virgo Collaboration) 2012
  {\em Phys. Rev. D\/} {\bf 85}(12) 122001

\bibitem{aLIGO}
{Harry G for the LIGO Scientific Collaboration} 2010 {\em Class. Quantum
  Grav.\/} {\bf 27} 084006

\bibitem{AdVirgo}
Acernese F {\em et~al.\/} (VIRGO Scientific) 2009 {\em Virgo Internal report:
  VIR–027A–09\/}
  \urlprefix\url{https://tds.ego-gw.it/itf/tds/file.php?callFile=VIR-0027A-09.pdf}

\bibitem{0264-9381-23-8-S26}
et~al B~W 2006 {\em Classical and Quantum Gravity\/} {\bf 23} S207

\bibitem{0264-9381-27-8-084004}
{K Kuroda and the LCGT Collaboration} 2010 {\em Classical and Quantum
  Gravity\/} {\bf 27} 084004

\bibitem{MCMCMethods}
{Christensen N and Meyer R} 1998 {\em Physical Review D\/} {\bf 58} 082001

\bibitem{NestedSampling}
{Veitch J and Vecchio A} 2010 {\em Physical Review D\/} {\bf 81} 062003

\bibitem{MultiNestLISA}
{Feroz F, Gair J, Graff P, Hobson M, and Lasenby A} 2010 {\em Class. Quantum
  Grav.\/} {\bf 27} 075010

\bibitem{MultiNestGW}
{Feroz F, Gair J, Hobson M, and Porter E} 2009 {\em Class. Quantum Grav.\/}
  {\bf 26} 215003

\bibitem{S6PE}
{Abadie J et al} (LIGO Scientific Collaboration and Virgo Collaboration) 2013
  {\em arXiv:1304.1775\/}

\bibitem{SMEE}
{Logue J et al} 2012 {\em Physical Review D\/} {\bf 86} 044023

\bibitem{BayesianReconstruction}
{Rover C et al} 2009 {\em Physical Review D\/} {\bf 80} 102004

\bibitem{0264-9381-27-17-173001}
{Abadie J et al} (LIGO Scientific Collaboration and Virgo Collaboration) 2010
  {\em Classical and Quantum Gravity\/} {\bf 27} 173001

\bibitem{X-Pipeline}
{Sutton P et al} 2010 {\em New Journal of Physics\/} {\bf 12} 053034

\bibitem{CoherentWaveBurst}
{Klimenko S, Yakushin I, Mercer A, and Mitselmakher G} 2008 {\em Class. Quant.
  Grav.\/} {\bf 25} 114029

\bibitem{STAMP}
{Thrane E et al} 2011 {\em Physical Review D\/} {\bf 83} 083004

\bibitem{Stochtrack}
{Thrane E and Coughlin M} 2013 {\em Phys. Rev. D\/} {\bf 88} 083010

\bibitem{Stochsky}
{Thrane E and Coughlin M} 2014 {\em Accepted to Phys. Rev. D\/}

\bibitem{Alford:2011pi}
{Alford M et al} 2012 {\em Physical Review D\/} {\bf D85} 044051

\bibitem{PhysRevD.58.084020}
{Owen B et al} 1998 {\em Physical Review D\/} {\bf 58}(8) 084020

\bibitem{burstegard}
Prestegard T and Thrane E 2012 {\em LIGO DCC\/}  L1200204
  \url{https://dcc.ligo.org/cgi-bin/DocDB/ShowDocument?docid=93146}

\bibitem{PhysRevD.85.042001}
{De Rosa R, Forte L, Garufi F, and Milano L} 2012 {\em Phys. Rev. D\/} {\bf
  85}(4) 042001

\bibitem{lrr-2009-2}
{Sathyaprakash B S and Schutz B} 2009 {\em Living Reviews in Relativity\/} {\bf
  12} \urlprefix\url{http://www.livingreviews.org/lrr-2009-2}

\bibitem{Skilling}
J S 2004 {\em AIP Conference Proceedings of the 24th International Workshop on
  Bayesian Inference and Maximum Entropy Methods in Science and Engineering\/}
  {\bf 735} 395--405
  \urlprefix\url{http://iopscience.iop.org/0264-9381/27/19/194002/}

\bibitem{MultiNest}
{Feroz F, Hobson MP, and Bridges M} 2009 {\em Mon. Not. Roy. Astron. Soc.\/}
  {\bf 398} 1601--1614

\bibitem{MatlabMultiNest}
{Pitkin M and Romano J} 2013 {\em
  http://ccpforge.cse.rl.ac.uk/gf/project/multinest/frs/\/}

\bibitem{PhysRevD.88.084044}
{Littenberg T et al} 2013 {\em Phys. Rev. D\/} {\bf 88}(8) 084044

\bibitem{PhysRevD.46.5236}
Finn L~S 1992 {\em Phys. Rev. D\/} {\bf 46}(12) 5236--5249
  \urlprefix\url{http://link.aps.org/doi/10.1103/PhysRevD.46.5236}

\bibitem{PhysRevD.49.2658}
Cutler C and Flanagan E~E 1994 {\em Phys. Rev. D\/} {\bf 49}(6) 2658--2697
  \urlprefix\url{http://link.aps.org/doi/10.1103/PhysRevD.49.2658}

\bibitem{PhysRevD.88.084013}
Rodriguez C~L, Farr B, Farr W~M and Mandel I 2013 {\em Phys. Rev. D\/} {\bf
  88}(8) 084013
  \urlprefix\url{http://link.aps.org/doi/10.1103/PhysRevD.88.084013}

\bibitem{durrett2010probability}
Durrett R 2010 {\em Probability: Theory and Examples\/} Cambridge series on
  statistical and probabilistic mathematics (Cambridge University Press) ISBN
  9781139491136

\end{thebibliography}

\appendix

\section{Distribution of cross-power SNR}
\label{sec:STAMPDist}

Cross-power signal-to-noise ratio $\rho$ is defined as
\begin{equation}
\rho \approx \frac{\mathrm{Re}(s_1(f) s_2(f))}{\sqrt{P1_{adj}(f) P2_{adj}(f)}}
\end{equation}
where $s_1(f)$ and $s_2(f)$ are Fourier transforms of times series from two detectors (hence complex numbers) and $P1_{adj}(f)$ and $P2_{adj}(f)$ are the (averaged) PSDs calculated from adjacent segments \cite{STAMP}. Up to a scaling factor, the numerator is known as Y, the signal estimate, and the denominator $\sigma_{Y}$, its error. $s_1(f)$ and $s_2(f)$ are each Gaussian variables with mean 0 and variance $\sigma^2$. The distribution of a new variable z defined as z = $s_1(f) s_2(f)$, known as a normal product, is given by the expression \cite{durrett2010probability}
\begin{equation} 
f(z) = \frac{1}{\sigma^2} K_{0} \left(\frac{|z|}{\sigma^2}\right)
\end{equation}
where $K_{0}(x)$ is the modified Bessel function of the second kind. As $s_1(f)$ and $s_2(f)$ are complex vectors, this is actually the sum of two normal products, which is known as a double exponential or Laplace distribution. This has a distribution of the form

\begin{equation}
f_{Y}(y) = \frac{1}{2\sigma^2} \mathrm{exp}{\left(\frac{-|y|}{\sigma^2}\right)}.
\end{equation}
This is the distribution of Y. The second step is to calculate the distribution of $\sigma_{Y}$. $P1_{adj}$ and $P2_{adj}$ are the average PSDs calculated from segments on either side of the segment used to calculate Y
\begin{equation}
P1_{adj} = \frac{1}{N} \sum_{j=1}^{N} P1_{j}, \hspace{6pt} P2_{adj} = \frac{1}{N} \sum_{j=1}^{N} P2_{j}.
\end{equation}
In the frequency domain, each $P1_{j}$ and $P2_{j}$ are chi-squared distributed variables, the sum of which have distributions of the form
\begin{equation}
f_{P}(z) = \frac{N^N}{2^N \sigma^{2N} \Gamma(N)} \mathrm{exp}{\left(\frac{N z}{2 \sigma^2}\right)} z^{N-1} \hspace{6pt} (z\geq0).
\end{equation}
The distribution for $\sigma_{Y}$ is then
\begin{equation}
f_{\sigma_{Y}}(z) = \frac{N^{2N}}{2^{2N-2} \sigma^{4N} (\Gamma(N))^2} K_0{\left(\frac{N z}{\sigma^2}\right)} z^{2N-1}.
\end{equation}
The final step is to combine the distributions for Y and $\sigma_{Y}$
\begin{equation}
f_{SNR}(z) = \frac{N^{2N}}{2^{2N-1} \sigma^{4N+2} (\Gamma(N))^2} \int_0^{\infty} |x| e^{- \left|\frac{x z}{\sigma^2}\right|} K_0{\left(\frac{N x}{\sigma^2}\right)} x^{2N-1} dx.
\label{eq:SNR}
\end{equation}

\begin{figure}[t]
 \includegraphics[width=4in]{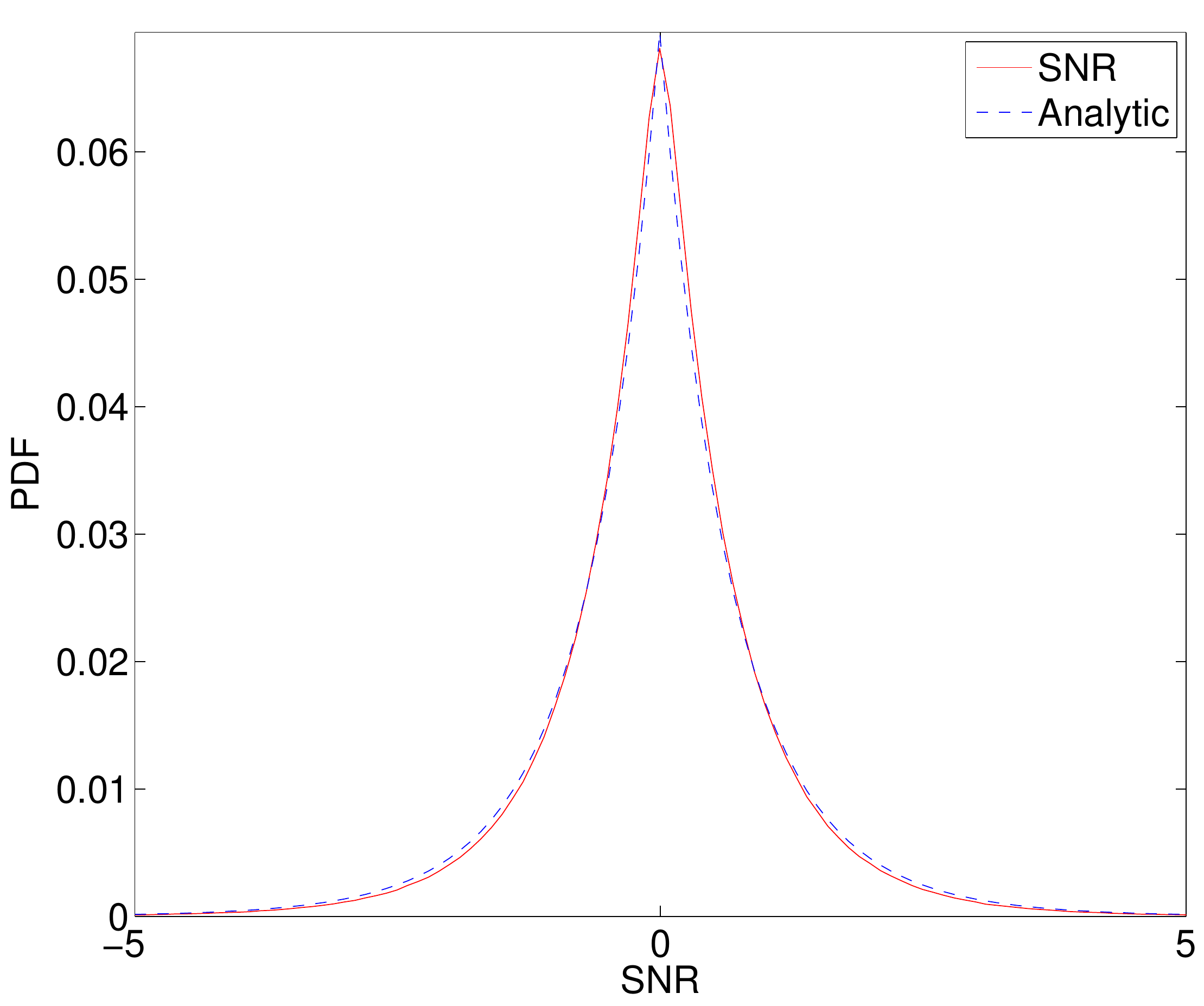}
 \caption{Probability density function of the pixels in a $ft$-map of cross-power SNR with the theoretical distribution given by Eq.~(\ref{eq:SNR}) overlaid.}
 \label{fig:STAMPHist}
\end{figure}

Fig. \ref{fig:STAMPHist} shows the probability density function (PDF) of the pixels of cross-power SNR used in this analysis overlaid with a distribution of $\rho(t;f)$ calculated from actual data. 

\section{Error approximation}
\label{sec:ErrorApproximation}

\begin{figure}[t]
 \includegraphics[width=4in]{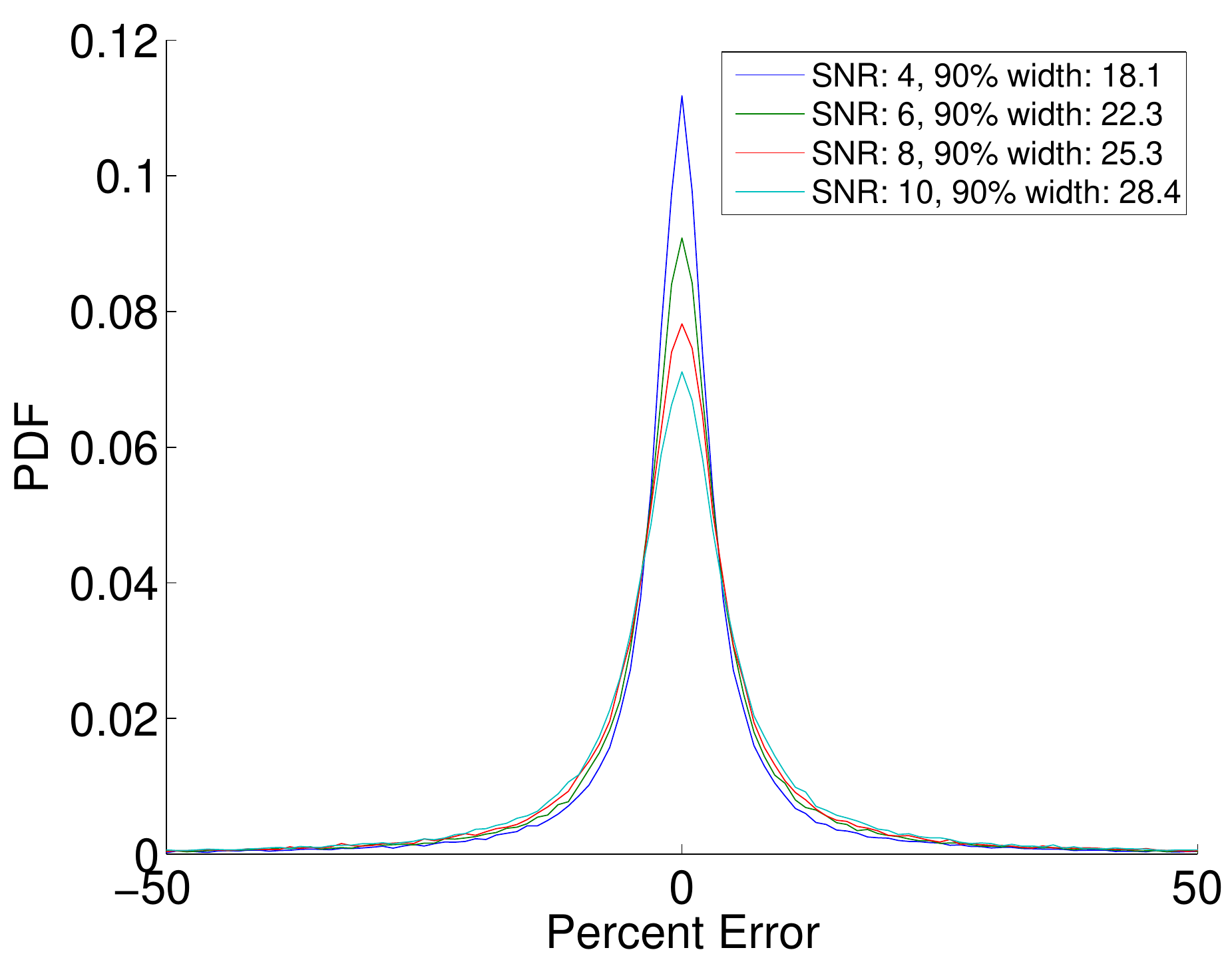}
 \caption{PDF of the percentage error of using $s_1 s_2 + n_1 n_2$ as an approximation for $h_1 h_2$. The width is the percentage error within which 90\% of the distribution is contained.}
 \label{fig:STAMPNoiseApproximation}
\end{figure}

If the timeseries of two detectors, $h_1$ and $h_2$, are composed of the sum of a signal and a noise part, (i.e., $h_1 = s_1 + n_1$ and $h_2 = s_2 + n_2$), when the two data streams are multiplied, the result will be in the form of

\begin{equation}
h_1 h_2 = s_1 s_2 + s_1 n_2 + s_2 n_1 + n_1 n_2
\label{eq:dataMultiple}
\end{equation}
The quantity $h_1*h_2$ is proportional to $\rho(t;f)$. The expectation values of the cross-terms, $s_1 n_2$ and $s_2 n_1$, are 0 because signal and noise are uncorrelated. To test the approximation that on average the cross-terms, $s_1 n_2$ and $s_2 n_1$, will sum to 0, sets of 100 pixels with different total SNRs associated with them are generated. The SNR for each set is computed by performing a sum of the individual pixel SNRs. This process seeks to imitate the total error accumulated due to the assumption above. In this case, the total error is the sum of $s_1 n_2$ and $s_2 n_1$ for all of the pixels.

Fig. \ref{fig:STAMPNoiseApproximation} shows the percent difference between $h_1 h_2$ and $s_1 s_2 + n_1 n_2$. For pixel sets with moderate total SNR, 90\% of the time, this approximation is within 25\% of its true value. Extremely high SNR events, which are an order of magnitude larger, yield errors on the order of 100\%. Examining the cross-terms, their contribution becomes more significant as the magnitude of the signal increases and the approximation breaks down in the high SNR regime. Conversely, the bias when using signals of moderate SNR is shown to be small in section \ref{sec:Results}.

\end{document}